\documentclass[preprint,12pt,3p]{elsarticle}
\usepackage{amssymb}
\usepackage{subcaption}

\begin{document}

\begin{frontmatter}

\title{Preparation of deuterated targets for nuclear astrophysical relevant reactions }

\author[label1]{Tanmoy Bar\corref{cor1}}
\address[label1]{Saha institute of Nuclear Physics, HBNI \\
1/AF Bidhannagar, Salt lake, Kolkata - 700064, INDIA
}
\ead{tanmoy.bar@saha.ac.in}
\author[label1]{Chinmay Basu}

\begin{abstract}
In this work, a simple and efficient thin (50-350 $\mu g/cm^2$)deuterated polyethylene target
preparation method is described. The method of easy removal of thin targets from casting
surface without using any cryogenic freezing has been described. The difference in thermal
expansion properties is used to separate films from glass slides. 3-$\alpha$
radioactive source is used to measure the target thickness. The material property of the
prepared targets is verified by Attenuated total reflection (ATR) method.

\end{abstract}


\end{frontmatter}
\section{Introduction }
In low energy nuclear astrophysics (d,p), (d,n), (d,t), (d,$\alpha$), (d, $^3He$) are very important deuteron induced reaction to study. These study will allow us to know nucleosynthesis \citep{ref1} better. For all these studies deuterium-containing targets are very essential. These targets are even more useful to study nuclear reactions in Trojan-Horse method framework \citep{ref2}\citep{ref3}. For decades deuterium has been used as projectile to study various nuclear astrophysics experiment. Now in inverse kinematics calculation scenario self supporting deuterium containing targets are essential. For last 50 years people are preparing these targets in there laboratory with various methods \citep{ref4}- \citep{ref9}. But most of them are used solvent casting methods to prepare the target. Recently Febbraro. et. al. \citep{ref9} showed preparation of these targets in details. They preferred cryogenic freezing for the removal of thin films ($< 100\mu g/cm^2)$ which is otherwise very difficult to remove. Here we have discussed more simplified and easy method of film removal from the casting surface. We have not used any sodium chloride kind of release agent in our glass slide as used by G. T. J Arnison \citep{ref7}. Because use these can cause non uniformity specially in case of thin foil preparation. We also discussed some more minor details one need to keep in mind while preparing films with such expensive materials. We also tried to give some idea about the amount needed to produce a particular thickness of film.
\section{Preparation Method}
For our preparation of deuteron target we have used deuterated polyethylene (commonly known as $CD_2$) which is obtained from the deuteration of polybutadiene –deuterated rich in 1,4 addition. Here we used solvent casting method to prepare our target. Now deuterated polyethylene (melting temperature $125^oC$) is soluble in few organic solvents whose boiling temperature is more than $CD_2$ such as xylene ($\sim 140^oC$), Cyclooctane ($\sim 150^oC$), Glycerol ($\sim 290^oC$) etc. While using Glycerol one must give extra care to the temperature because its boiling temperature is much higher than the $CD_2$ material and overheating too much will turn polyethylene into gel. \\
\subsection{Solution preparation}
Here we have used p-xylene (any other isomer o- or m-xylene or mixture of all will also do that job) as our solvent to prepare the solution. 5ml of p-xylene is used to prepare the solution which is sufficient to cover 4-5 micro slides (each having dimension 1 inch $\times$ 3 inch) convincingly. We need to ensure minimal wastage of material during this preparation and take extra care form being contaminated by chemical or dust particles. For that reason we have prepared our film in controlled environment clean lab to ensure proper temperature, humidity and clean environment during the preparation. Controlled temperature is very important to ensure while handling xylene kind of chemical which is flammable in nature above the surrounding temperature above $25^oC$. So it is recommended to perform the entire process inside a enclosure with a exhaust fan on top to eliminate any sort of health hazards by xylene \citep{ref10}.
For the solution preparation purpose we have used a beaker as small as possible to ensure the minimum waste due to the sticking of material on the surface of the beaker. For the solution preparation purpose we used double boiler[Figure 1] technique as described in \citep{ref9}. Here maintaining the temperature in the upper beaker is very crucial, we just need to boil the lower beaker containing only p-xylene and vapour generated from it will transfer sufficient heat to the upper beaker which was covered with aluminium foil to dissolve $CD_2$ material completely. Now solution was heated for almost an hour or until it was visually clear. It is always recommended to heat it for 10-15 minutes extra after solution became visually clear to ensure there is no residue left. Now, we also need to ensure that we do not go out of p-xylene in the bottom container before the complete dissolution of $CD_2$ material in the top beaker. We used a typical chemistry glass beaker of 50 mL volume and 40 mL of p-xylene was enough to ensure more than 1 hour of boiling without any issue.\\
\begin{figure}[!h]
\begin{center}
\includegraphics[scale=.5]{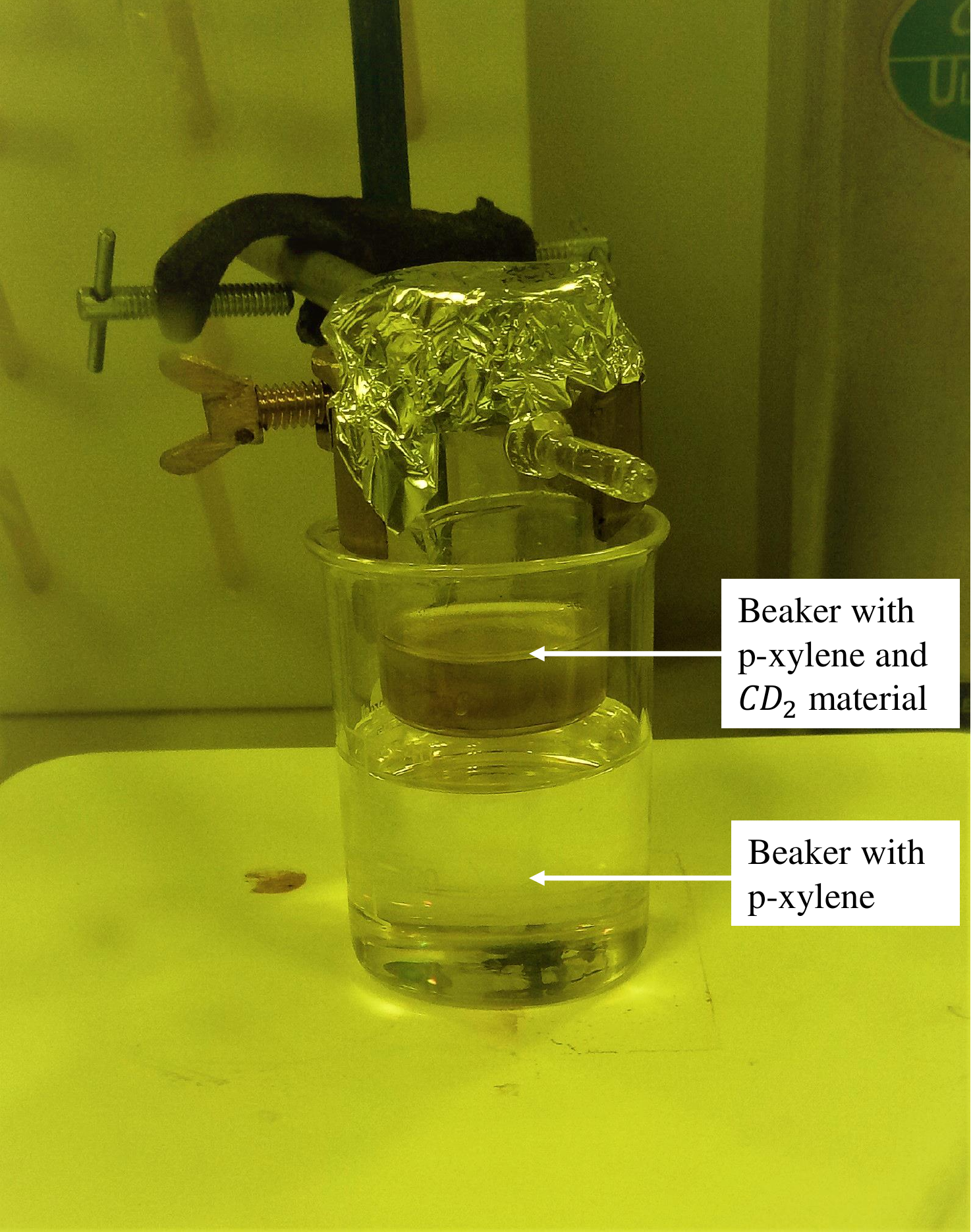}
\caption{Setup for the solution preparation}
\end{center}
\end{figure}
\subsection{Casting on glass slides}
After all the material is dissolved properly we took a clean thin glass stick to give mixture a quick whisk to ensure a homogeneous mixture before pouring onto the glass slides. Then solution was poured onto four glass slides which were placed side by side on a levelled surface. This is one of crucial and need to do it quickly otherwise $CD_2$ material inside solution will start to precipitate and we will have a non-uniform film. After that slides were left to air for 5-10 mins. so that xylene can evaporate from the solution and only $CD_2$ is left.\\
\subsection{Removal of films}
Now removal of those films from the casting surface is quite difficult specially for thinner targets ($50-100\mu g/cm^2$).
After that glass slides are placed on the hot plate ($>140^oC$) for 30-60 sec. Then those slides were carefully emerged inside a large water beaker at once with the help of a tweezers directly from hot plate. After few seconds films were removed from the slides mechanically by a sharp tweezers. Due to different thermal expansion of glass slides and polyethylene it will be much more easier to remove the film from the slides. We have done our entire process of film removal keeping slides under water. By this techniques we have prepared several thin targets($50-350\mu g/cm^2$) successfully[Figure 2 and Figure 3].
\begin{figure}
  \centering
  \begin{subfigure}[b]{0.48\linewidth}
    \includegraphics[width=\linewidth]{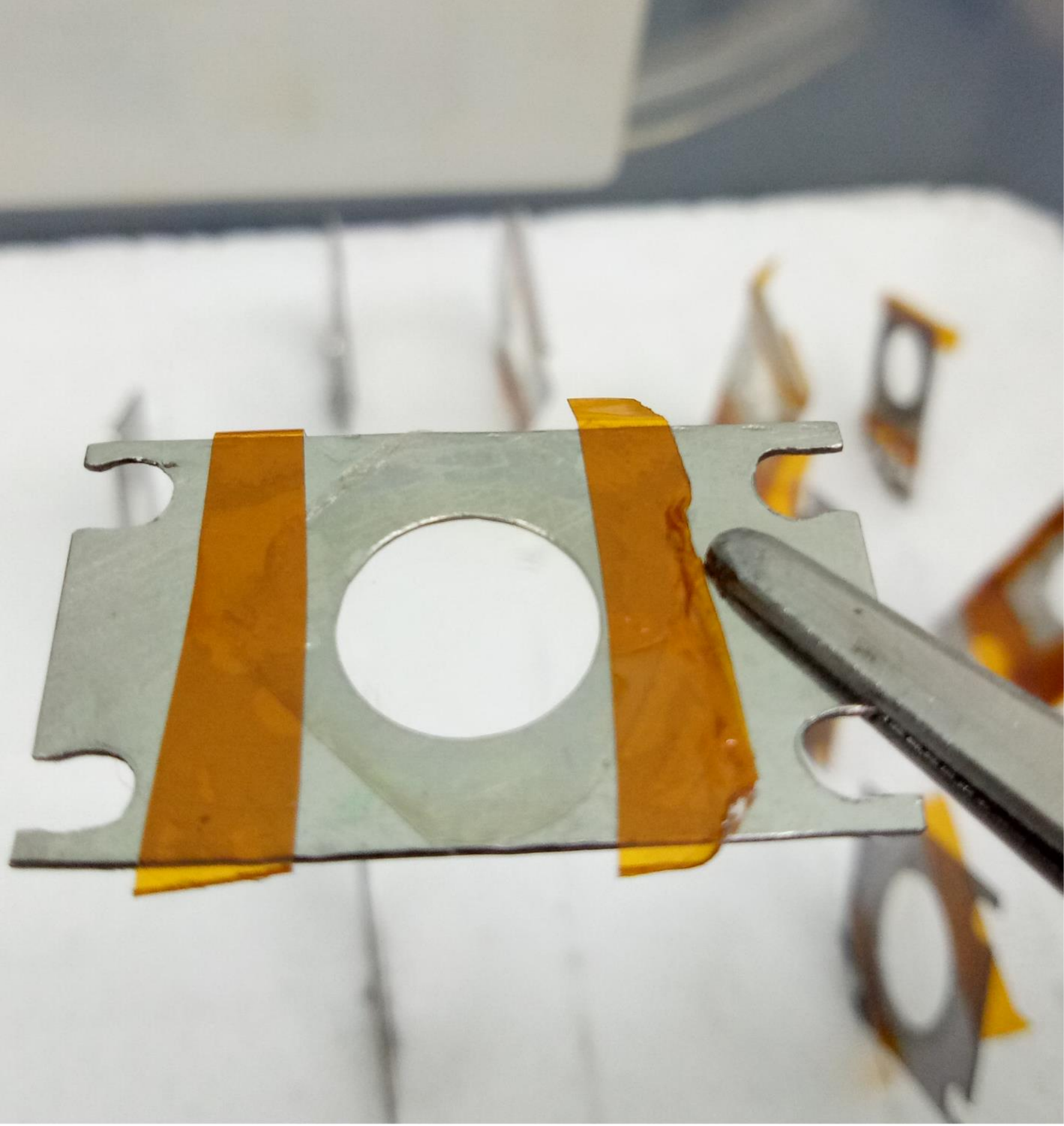}
    \caption{$CD_2$ target with thickness $\sim 280 \mu g/cm^2$}
  \end{subfigure}
  \begin{subfigure}[b]{0.48\linewidth}
    \includegraphics[width=\linewidth]{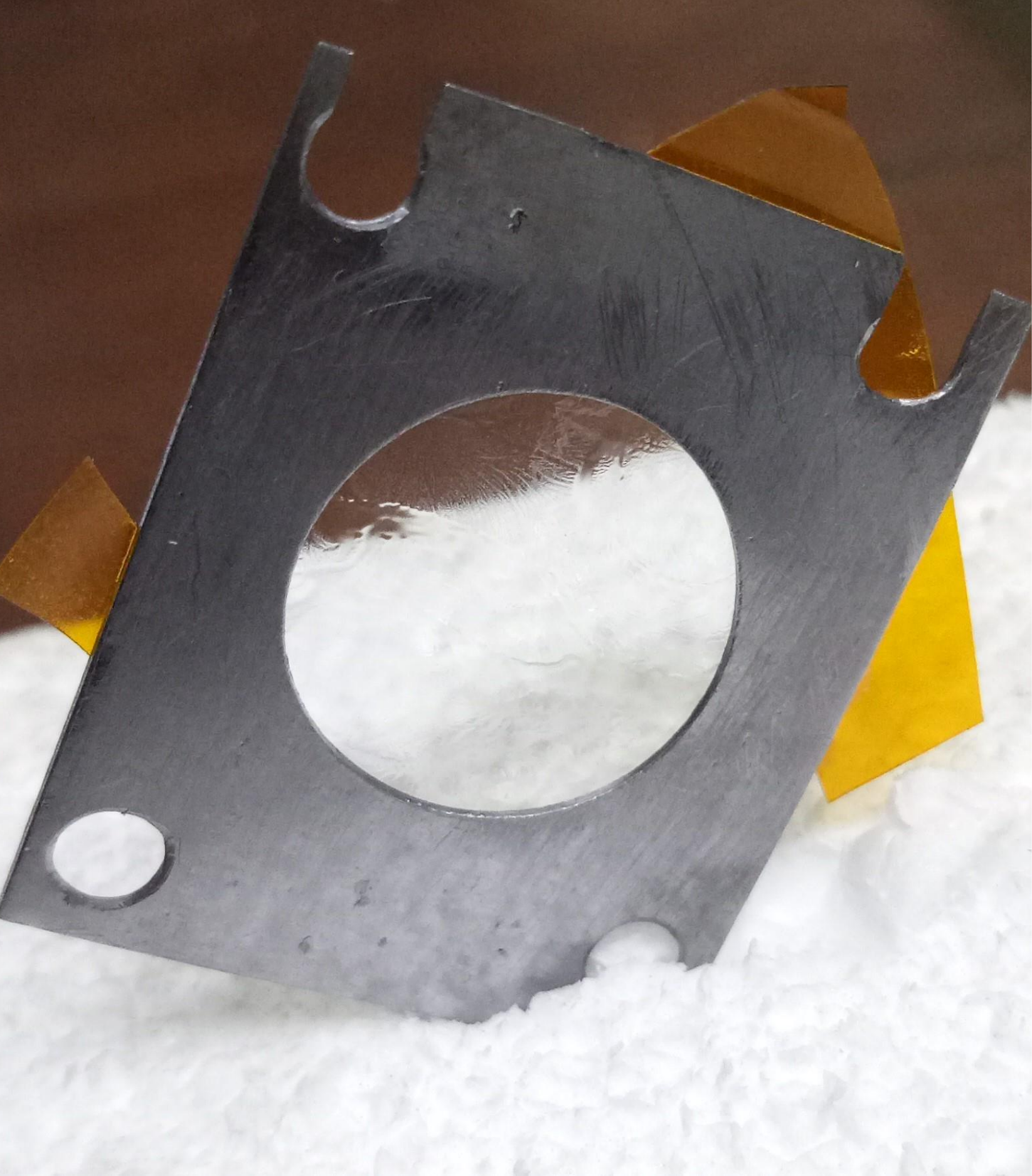}
    \caption{$CD_2$ target with thickness $\sim 50 \mu g/cm^2$}
  \end{subfigure}
  \caption{Deuterated polyethylene targets mounted on frame.}
\end{figure} 
\section{Analysis of prepared films}
For the thickness measurement purpose we have used 3-$\alpha$ source and energy loss simulation software SRIM \citep{ref11}. Now during the preparation target material is mixed with chemicals, it was heated for several times and lastly emerged inside water before removing it mechanically. It is very important to make sure there is no chemical degradation or contamination happened during the preparation process. To ensure this we have done Attenuated total reflection(ATR) test to our prepared foils. These result will show about the condition of C-H and C-D bonding inside the film. Figure 4 is one of the ATR results of our prepared foils. Variation of absorbance with wave number will represent different bonding of elements presents inside the film.

\begin{figure}[!h]
\begin{center}
\includegraphics[scale=.5]{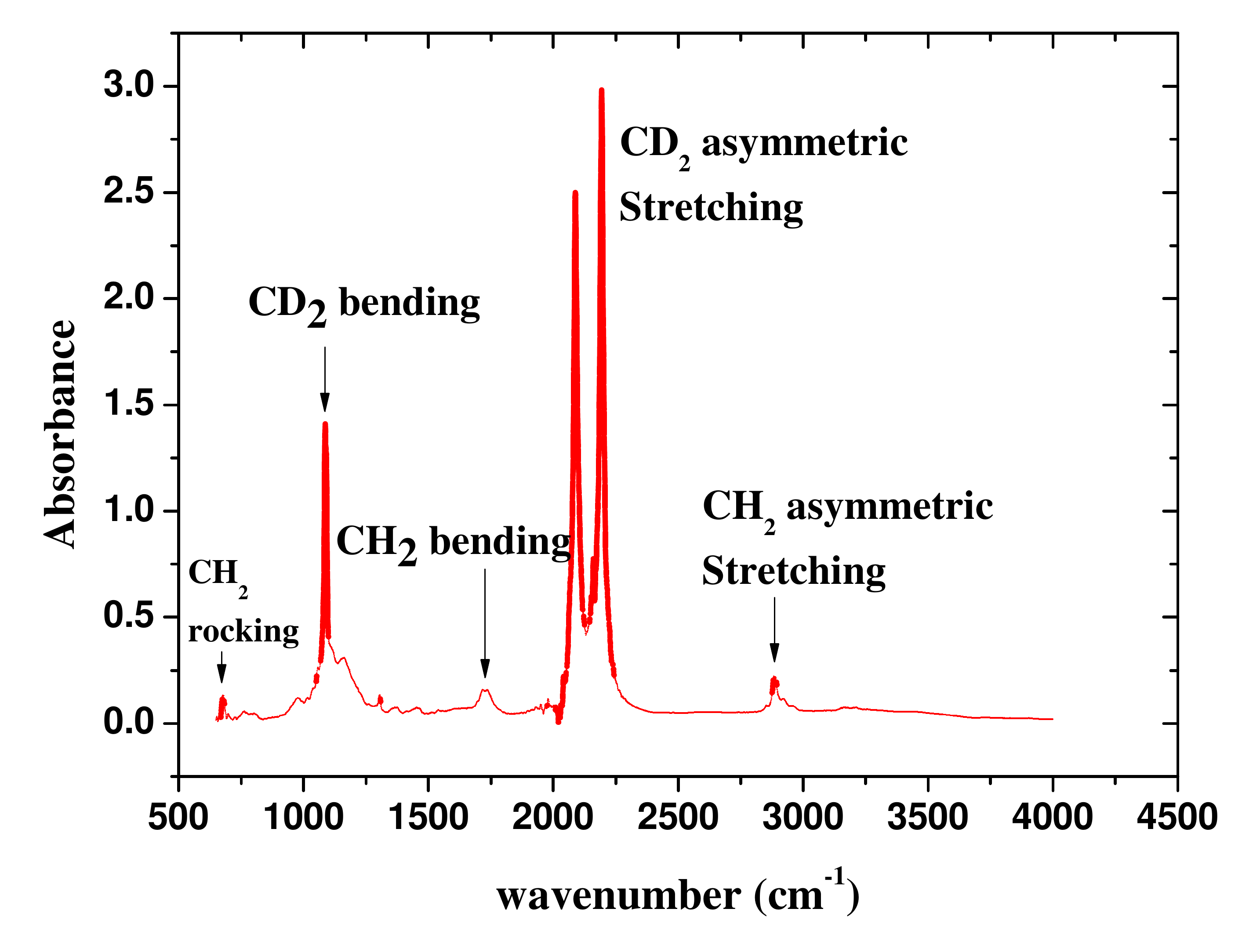}
\caption{ATR result of sample which shows different bonding of elements.}
\end{center}
\end{figure}

\section{Summary and discussion }
   Here we have shown very easy and efficient way of preparing deuterated polyethylene targets without any need of rapid cryogenic freezing facility even for thin targets. Now amount of material needed to prepare a film of specific thickness is very tricky. It is highly depend on the volume of beaker one is using to prepare the solution. Because the amount of loss due to sticking on the walls of that container will give great effect on prepared thickness. Pouring of solution is very important, uniformity in pouring will also ensure uniformity in prepared targets. Our prepared foil is little ivory in color, this coloration is because of the trace amount (finger prints) of the Wilkinson catalyst used while preparing the target material itself. By the ATR results we ensured the quality of material after the preparation of films. 
\section*{Acknowledgement}
\bibliographystyle{elsarticle-num}

\begin{thebibliography}{9}
\bibitem{ref1} C. Rolf, W. Rodney, Cauldron in the Cosmos: Nuclear Astrophysics, 1988.
\bibitem{ref2} C. Spitaleri et. al. PHYSICAL REVIEW C, VOLUME 60, 055802
\bibitem{ref3} M. La Cognata et. al. The Astrophysical Journal Letters, 739:L54 (6pp), 2011 October 1
\bibitem{ref4} C. M. Bartle et. al. NUCLEAR INSTRUMENTS AND METHODS 112 (1973) 615
\bibitem{ref5} C. M. Bartle et. al. NUCLEAR INSTRUMENTS AND METHODS 144 (1977) 599
\bibitem{ref6} G. E. Tripard et. al. Review of Scientific Instruments 38, 435 (1967)
\bibitem{ref7} G. T. J ARNISON NUCLEAR INSTRUMENTS AND METHODS 40 (I966) 359
\bibitem{ref8} M. A. OLIVO and G. M. BAILEY NUCLEAR INSTRUMENTS AND METHODS 57 (I967) 353-354
\bibitem{ref9} M. Febbraro et. al. Nuclear Instruments and Methods in Physics Research B 410 (2017) 53–59
\bibitem{ref10} Reena Kandyala et. al. J Oral Maxillofac Pathol. 2010 Jan-Jun; 14(1): 1–5.
\bibitem{ref11}SRIM--The stopping and range of ions in matter (2010)Nuclear Instruments and Methods in Physics Research Section B: Beam Interactions with Materials and Atoms 268(2010) 1818-1823.

\end{thebibliography}

\end{document}